\documentclass[3p,times]{elsarticle}
\usepackage{ecrc}

\newcommand{\avg}[1]{\left\langle #1 \right\rangle}

\volume{00}

\firstpage{1}

\journalname{} 

\runauth{G. Soyez}

\jid{procs}

\jnltitlelogo{Procedia Computer Science}

\usepackage{amssymb}

\usepackage[figuresright]{rotating}

\begin{document}

\begin{frontmatter}


\title{Seeing jets in heavy-ion collisions}


\author[ipht]{Gregory Soyez}
\address[ipht]{Institut de Physique Th\'eorique, CEA Saclay, CNRS URA 2306, F-91191 Gif-sur-Yvette, France}

\begin{abstract}
  In these proceedings, we briefly review how jets can be
  reconstructed in heavy-ion collisions. The main point we address is
  the subtraction of the large contamination from the underlying event
  background. We first present the main ingredients needed to define
  the jets and perform the background subtraction and then discuss the
  efficiency of the subtraction for different jet algorithms and
  background-estimation methods. 
\end{abstract}

\end{frontmatter}


\section{Introduction}\label{sec:intro}

Jets, naively seen as a proxy for the hard partons produced in a
final state, have always been powerful tools in collider physics. 
Recently, interest has grown in using jets in heavy-ion (HI) collisions as
a probe of the properties of the dense medium created in these collisions.

The main limitation to reconstruct jets in HI collisions is the fact
that they have to be isolated from the large Underlying Event (UE), of
${\cal O}(100-300)$ GeV per unit of $y-\phi$ area. 
This UE contamination to the jet has to be subtracted if one wants to
reconstruct its momentum correctly and obtain an unbiased measurement
of the effects of the medium.

In practice, both RHIC and the LHC have already accumulated a large
amount of data and have tried various approaches to address that issue
(see {\em e.g.} \cite{star1,star2,phenix,atlas}).

Here, we describe a jet-based method \cite{hipaper} for full jet
reconstruction derived from the original proposal \cite{bkgestim} for
pileup subtraction in $pp$ collisions. 
We shall first present the method (see eq. (\ref{eq:subtraction})
below) and describe its main elements: defining a jet, defining its
area and estimating the background density per unit area. 
We shall then study using Monte-Carlo simulations what kind of performances
one might expect in practice.

Most of the effects we shall describe below can be estimated
analytically. However, since this often involves rather technical
arguments, we will not reproduce them in these proceedings and
redirect the interested reader to Ref. \cite{hipaper}.
Furthermore, we shall concentrate on the situation at the LHC. Similar
results are obtained for RHIC and can be found in \cite{hipaper}.

\section{Defining background-subtracted jets}\label{sec:framework}

\paragraph{Master formula} 

If the background one wants to subtract from the jets is sufficiently
uniform, one can characterise it by two numbers: its average
transverse-momentum density per unit area $\rho$ and its fluctuations
per unit area $\sigma$. For a given jet, the contamination due to the
additional background will thus be $\rho A_{\rm jet}\pm \sigma
\sqrt{A_{\rm jet}}$, where $A_{\rm jet}$ is a measurement of the area
of that jet. To correct for the background contamination, we can thus
define a (4-vector) {\em subtracted jet} momentum as
\begin{equation}\label{eq:subtraction}
p_j^{\mu, {\rm sub}} = p_j^\mu - \rho A_j^\mu \, .
\end{equation}
In order to apply that master formula, three building blocks,
described below, are necessary: define jets, define their area and
estimate the background density $\rho$.

Before getting to that, let us mention that eq. (\ref{eq:subtraction})
is incomplete in two major aspects. The first one comes
from the fluctuations of the background, leaving an uncertainty of the
order of $\sigma \sqrt{A_{\rm jet}}$. The second source is referred to
as {\em back reaction} \cite{areas}: since background particles have a
non-vanishing momentum, they can affect the clustering of the ``hard''
particles; this means, on top of the pure background contamination
($\rho A_{\rm jet}$), the hard contents of the jet can change when
considering it together with the background. We will show some
examples of these effects in Section~\ref{sec:simulations}.

\paragraph{Jet definitions}

Defining jets from a list of input objects --- {\em e.g.} particles or
calorimeter towers --- has been the topic of many discussions over the
last few decades (see {\em e.g.} \cite{gavinsreview}). Here, we shall
concentrate on recombination-type algorithms that define the distances
\begin{eqnarray}
d_{ij} & = & {\rm min}(k_{t,i}^{2p},k_{t,j}^{2p})
             \left(\Delta\phi_{ij}^2 + \Delta y_{ij}^2\right),\\
d_{iB} & = & k_{t,i}^{2p} R^2,
\end{eqnarray}
respectively between any pair of particles and for a single
particle. The clustering works by successively identifying the
smallest distance; if it is a $d_{ij}$ recombine particles $i$ and
$j$, otherwise, call $i$ a jet.

The geometric factor in $d_{ij}$ basically ensures that collinear
particles will be clustered in the same jet. The $k_t$-dependent
prefactor varies from one algorithm to another: the $k_t$ algorithm
\cite{kt1,kt2,kt3} has $p=1$ so that soft QCD emissions are clustered
early in the sequence; the simple case $p=0$ corresponds to the
Cambridge/Aachen (C/A) algorithm \cite{cam1,cam2}, useful {\em e.g.}
for jet substructure studies; the anti-$k_t$ algorithm \cite{antikt},
$p=-1$, will cluster jets around hard objects, producing hard jets
with rigid, circular, shapes. Note that the anti-$k_t$ algorithm is
the default choice for most of the LHC experiments. All these
algorithms are accessible using the FastJet package
\cite{fastjet,fastjet_web} and we shall fix $R=0.4$ in what follows.

Recently, a lot of progress has been made that involve making use of
the substructure of the jets. To illustrate that, we shall consider,
on top of the 3 algorithms introduced above, the C/A algorithm
supplemented with a filter (C/A(filt)) \cite{boosted_higgs}. The
latter works by reclustering each individual jet with a smaller radius
(we shall use $R/2$) and only keep the 2 hardest subjets. Because of
the collinear nature of QCD branchings, the filtering procedure is
expected to keep most of the QCD part of the jet and reject part of
the Underlying-Event contamination.

\paragraph{Jet areas}

Practically, the area of a jet is defined \cite{areas} as the region
in which it captures (infinitely) soft particles (ghosts). Here, we
will use the {\em active} area of jets determined by adding a dense
coverage of ghosts to the event\footnote{This will not modify the
  clustering of the hard jets, provided the algorithm is
  infrared-safe.}, each carrying a ``quantum'' of area, clustering
them together with the jets, and obtain the area by summing the ghost
it contains. This procedures mimics the addition of a background to
the event with the exception that ghosts are infinitely soft.

The precise details of how the jet area is defined are largely
irrelevant. However, we would like to stress that for the computation
of the background density (see below), we advise the use of the {\sf
  explicit\_ghost} option of FastJet \cite{fastjet} to correctly handle
empty regions of the event.

\paragraph{Background estimation}

In \cite{bkgestim}, the following method has been suggested: first
cluster the event with the $k_t$ or C/A algorithm with a
radius\footnote{This choice of jet definition for the estimation of
  the background is independent of the choice made to reconstruct the
  jets in the event. For the former, we suggest using the $k_t$ or C/A
  algorithm as it avoids having many jets with a small area.} $R_\rho$
to use, as an estimator for the background density per unit area,
\begin{equation}\label{eq:median}
  \rho = \mathrm{median}\left\{\frac{p_{j,t}}{A_{j,t}}\right\},
\end{equation}
where the computation of the median includes all the jets up to a
maximal rapidity $y_{\rm max}$. We refer to that choice as the {\em
  global} range below.

To better take into account the non-uniformities of the background, we
have realised that there is some interest in using only jets in the
vicinity of the jet we want to subtract:
\begin{equation}\label{eq:local_median}
  \rho_{{\cal R(j)}} = \mathrm{median}\left\{\frac{p_{j',t}}{A_{j',t}}\right\}_{j' \in {\cal R(j)}},
\end{equation}
where the subscript ${\cal R(j)}$ explicitly refers to the fact that
we use a {\em local range} centred on the jet $j$ for which we want to
estimate the background\footnote{Note that if the use of a local range
  allows to better estimate of the background at the position of the
  jet we want to subtract, it may also have the inconvenient to
  increase the sensitivity to quenching effects.}.

We have considered 3 different options: a {\em
  CircularRange}($\Delta$) keeping jets within a distance $\Delta$ (we
used $\Delta=3R$) of the jet, a {\em DoughnutRange}($\delta,\Delta$)
keeping jets with a distance from the jet between $\delta$ and
$\Delta$ (we used $\delta=R$, and $\Delta=2R$ or $3R$) and a {\em
  StripRange}($\Delta$) being a rapidity strip extending between
$y_{\rm jet}-\Delta$ and $y_{\rm jet}+\Delta$ (we used $\Delta=2R$ or
$3R$).

The use of a median in (\ref{eq:median}) and (\ref{eq:local_median})
rather than an average is made to reduce the effect of the hard jets
on the estimation of $\rho$. 
In order to further reduce that bias, we have also tested the exclusion
of the two hardest jets in the event from the jets used for the median
computation.

\section{Expected performances}\label{sec:simulations}

\paragraph{Framework}

\begin{figure}
\begin{center}
\includegraphics[angle=270,width=0.9\textwidth]{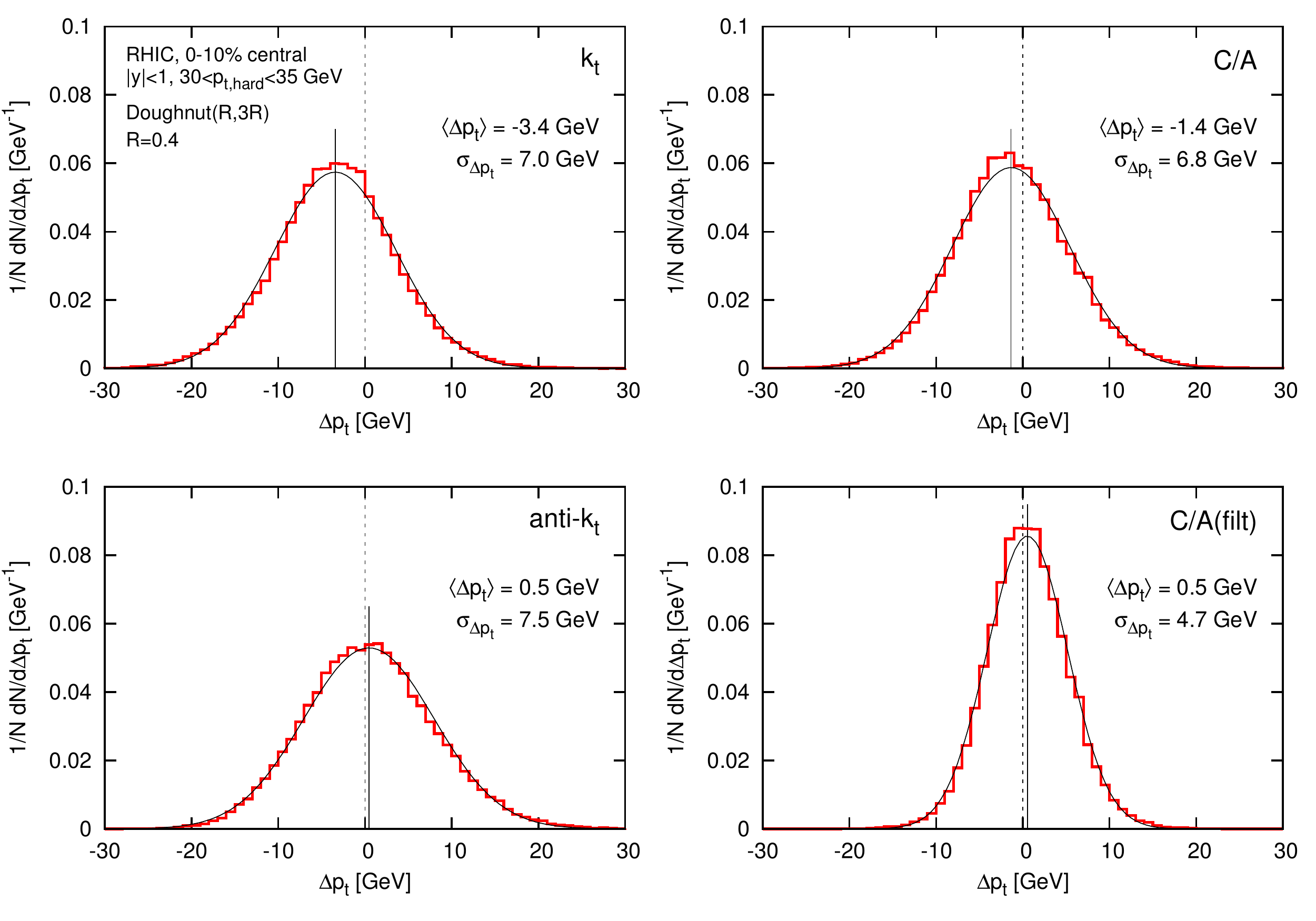}
\end{center}
\caption{Distribution of $\Delta p_t$ (red histograms) at RHIC for
  each of our 4 jet algorithms, together with a Gaussian (black curve)
  whose mean (solid vertical line) and dispersion are equal to
  $\avg{\Delta p_t}$ and $\sigma_{\Delta p_t}$
  respectively.}\label{fig:histo}
\end{figure}

In this Section, we study how the method presented above performs in
practice. 

To achieve that, we shall embed $pp$ hard events in HI background,
apply our background-subtraction method and see how precisely the
original hard jets are reconstructed. The $pp$ events will be
generated using Pythia 6.4 \cite{pythia} and for the $AA$ events we
have used Hydjet 1.6 \cite{hydjet}. We note that, for 0-10\% central
collisions, the HI background has an average density $\langle \rho
\rangle$ of about 100 GeV at RHIC and 310 GeV at the LHC, with
fluctuations $\langle \sigma \rangle$ around 8 and 20 GeV, respectively.

Jets are reconstructed and subtracted for both the hard event alone
and the hard event embedded in the heavy-ion background (which we
shall refer to as the full event). We have concentrated on the two
hardest jets in the hard event. For each of them, we search for a
matching full jet in the full event by requiring that the hard
contents of the full jet accounts for at least 50\% of the original
hard jet\footnote{Though we do not show it here, the matching
  efficiencies depend slightly on the jet definition and are typically
  above 95\% for $p_t>10$ GeV at RHIC and above 98\% for $p_t>40$ GeV
  at the LHC.}.

For a matching pair of jets, we consider the difference between the
subtracted full jet and the subtracted hard jet:
\begin{equation}\label{eq:deltapt}
\Delta p_t = p_t^{AA,sub} - p_t^{pp,sub}.
\end{equation}

An efficient background subtraction would directly translate into
small $\Delta p_t$. Though we could obtain the full distribution of
$\Delta p_t$, see Fig. \ref{fig:histo} for RHIC kinematics, they can
be considered as sufficiently Gaussian for the purpose of comparing
different background-estimation ranges and jet-reconstruction
algorithms, and we shall concentrate on its average $\avg{\Delta p_t}$
and dispersion
\[
\sigma_{\Delta p_t} = \sqrt{\avg{\Delta p_t^2} - \avg{\Delta p_t}^2}.
\]
An effective subtraction would then be characterised by small values
of $\avg{\Delta p_t}$ and $\sigma_{\Delta p_t}$.

For the results presented below, the $k_t$ algorithm with a radius
$R_\rho=0.5$ has been used for the median-based estimation of the
background density.

\paragraph{Choice of range}

\begin{figure}
\begin{center}
\includegraphics[angle=270,width=0.45\textwidth]{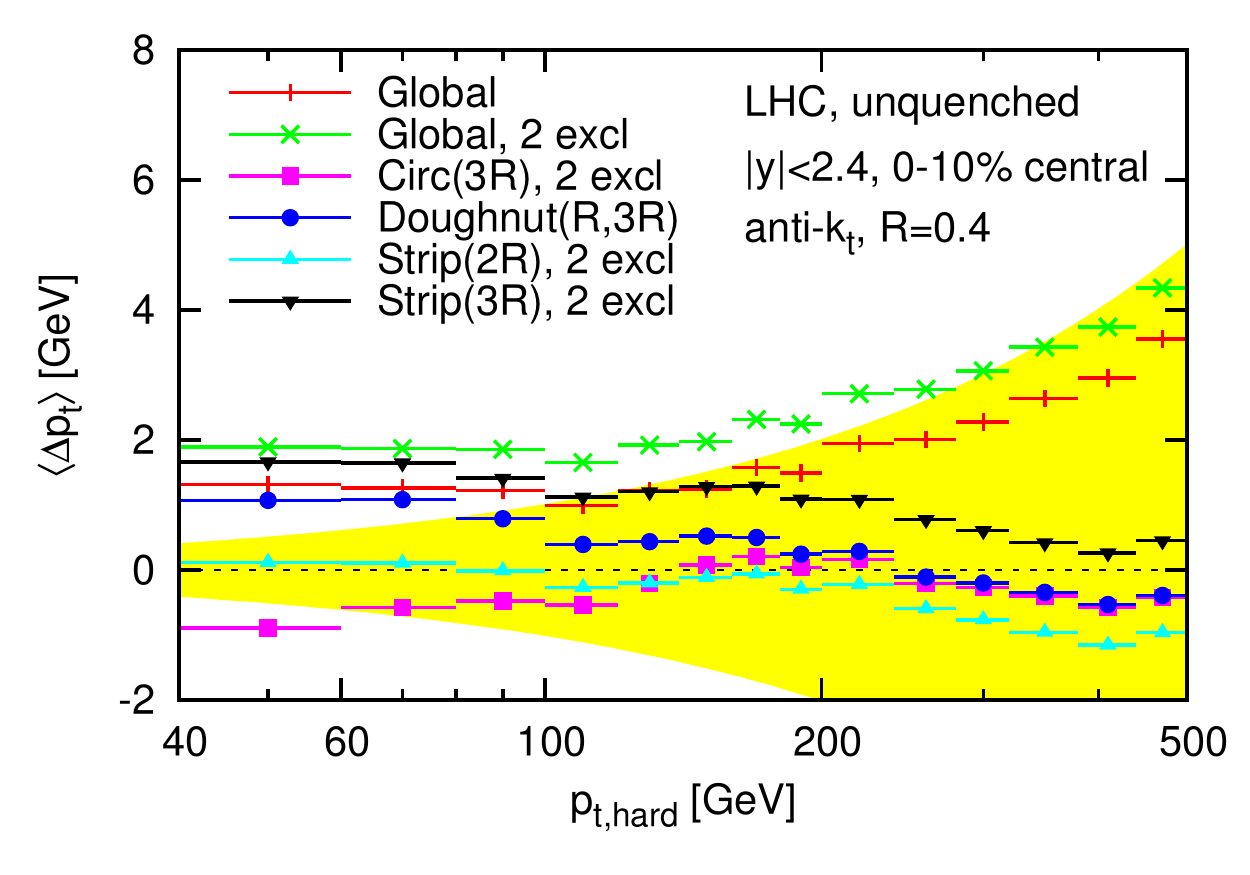}
\end{center}
\caption{Effect of the choice of range on the average $p_t$ shift,
  $\Delta p_t$, as defined in eq.~(\ref{eq:deltapt}), for LHC
  kinematics. In this figure and those that follow, the yellow band
  corresponds to 1\% of the $p_t$ of the hard jet. The label ``2
  excl'' means that the two hardest jets of the event have been
  excluded from the median computation.}\label{fig:ranges}
\end{figure}

We start our investigation by looking at how the choice of the range
made to estimate the background density affect the subtraction. On
Fig. \ref{fig:ranges}, we see that for all the ranges we consider, the
average uncertainty on the reconstructed jet transverse momentum is
$\lesssim$ 1\%, with a small preference\footnote{For a limited
  rapidity coverage, a global range behaves as a local
  range. However, at large $p_t$, jets tends to be produced at smaller
  rapidity where the background is larger and a global estimation of
  $\rho$ leads to an under-subtraction.} for local ranges at large
$p_t$.

In choosing the range, there is a tension between remaining in the
vicinity of the jet one wants to subtract and having enough jets in
the range to estimate the median properly. As a rule of thumb, it can
be shown that a local range needs to contain at least 9 jets to give a
reasonable estimate of $\rho$.

For what follows, we shall use the DoughnutRange($R,3R$) as a default choice.

\paragraph{Choice of algorithm}

\begin{figure}
\begin{center}
\includegraphics[angle=270,width=0.45\textwidth]{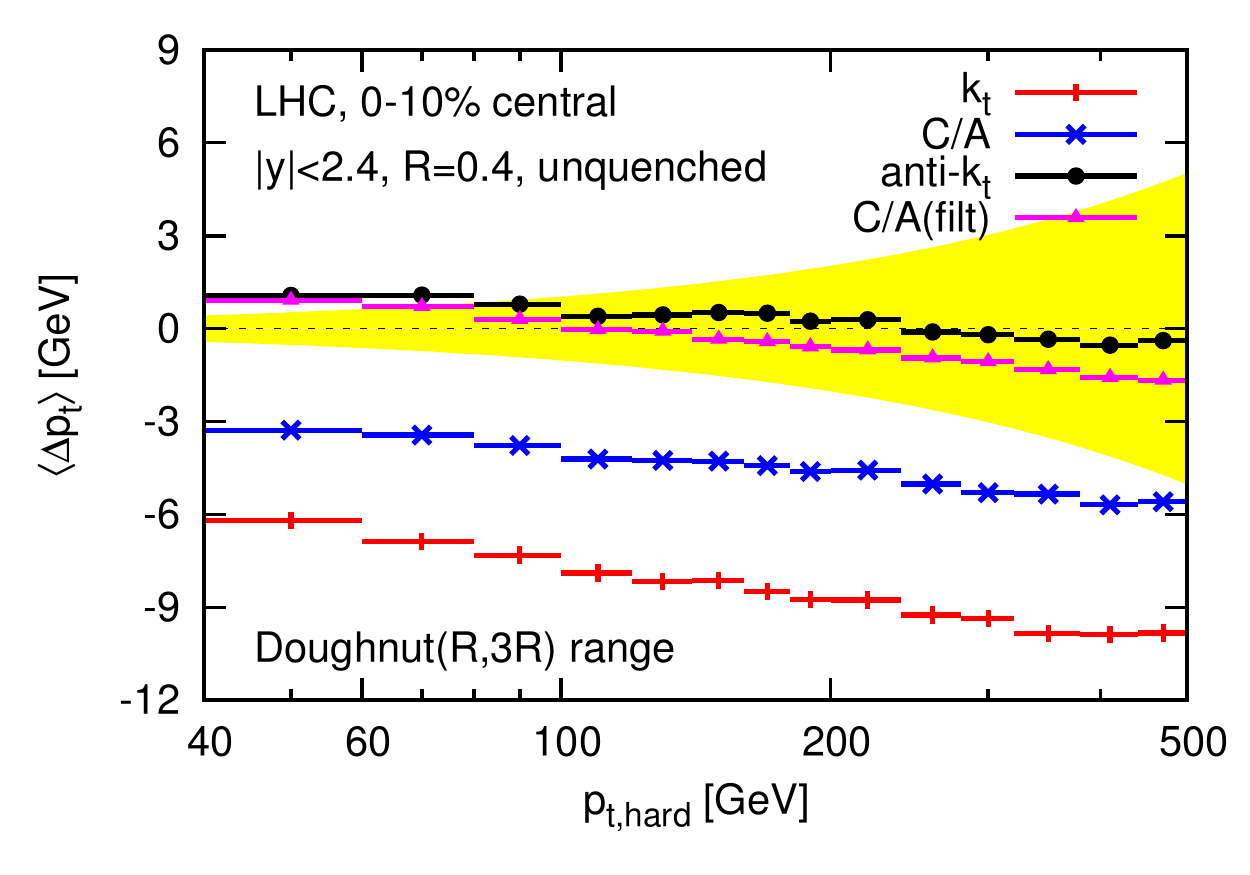}
\includegraphics[angle=270,width=0.45\textwidth]{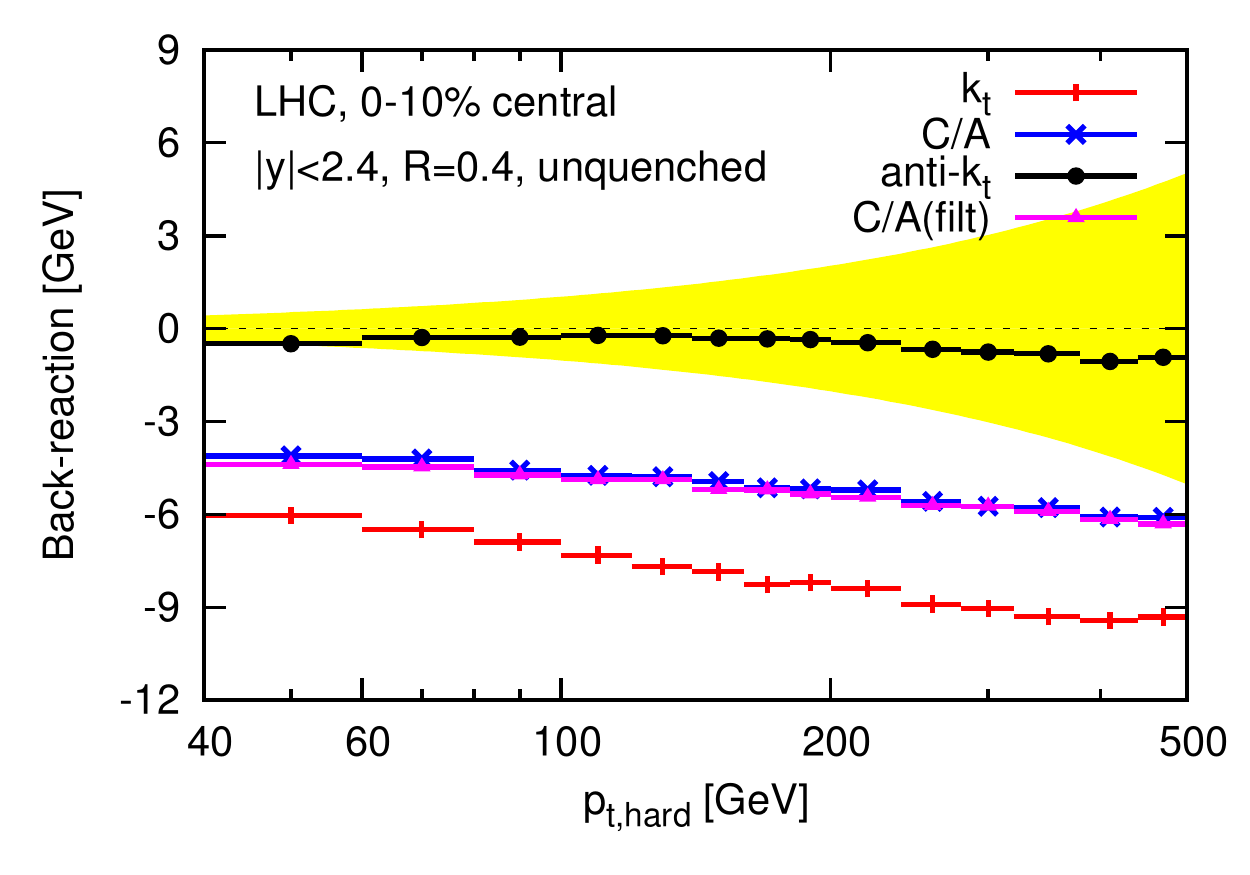}
\end{center}
\caption{Left: average shift $\avg{\Delta p_t}$, as a function of
  $p_{t,\rm hard}$, shown for the LHC. Right: contribution to
  $\avg{\Delta p_t}$ due to back-reaction. Note that the back-reaction
  results are independent of the range used for estimating $\rho$ in
  the heavy-ion event.}\label{fig:shift}
\end{figure}

We now focus on the dependence on the specific choice of algorithm and
first consider the average shift $\avg{\Delta p_t}$. It is plotted on
Fig. \ref{fig:shift}(left) for the four algorithms under
consideration. We see that while the anti-$k_t$ and C/A(filt)
algorithms give an average shift close to zero, the $k_t$ and C/A
algorithms have a larger, negative, shift. 

Since the estimation of the background density $\rho$ is common to all
4 algorithms, this can actually be traced back to a difference in
back-reaction as seen on Fig. \ref{fig:shift} (right). For all except
C/A(filt), the differences observed in the average shift correspond to
differences in back-reaction and the fact that the anti-$k_t$ has a
small shift is a direct consequence of its rigidity. In the case of
the C/A(filt) the final small shift can be explained by a (fortuitous)
cancellation between the offset due to back-reaction and a filtering
bias.

\begin{figure}
\begin{center}
\includegraphics[angle=270,width=0.45\textwidth]{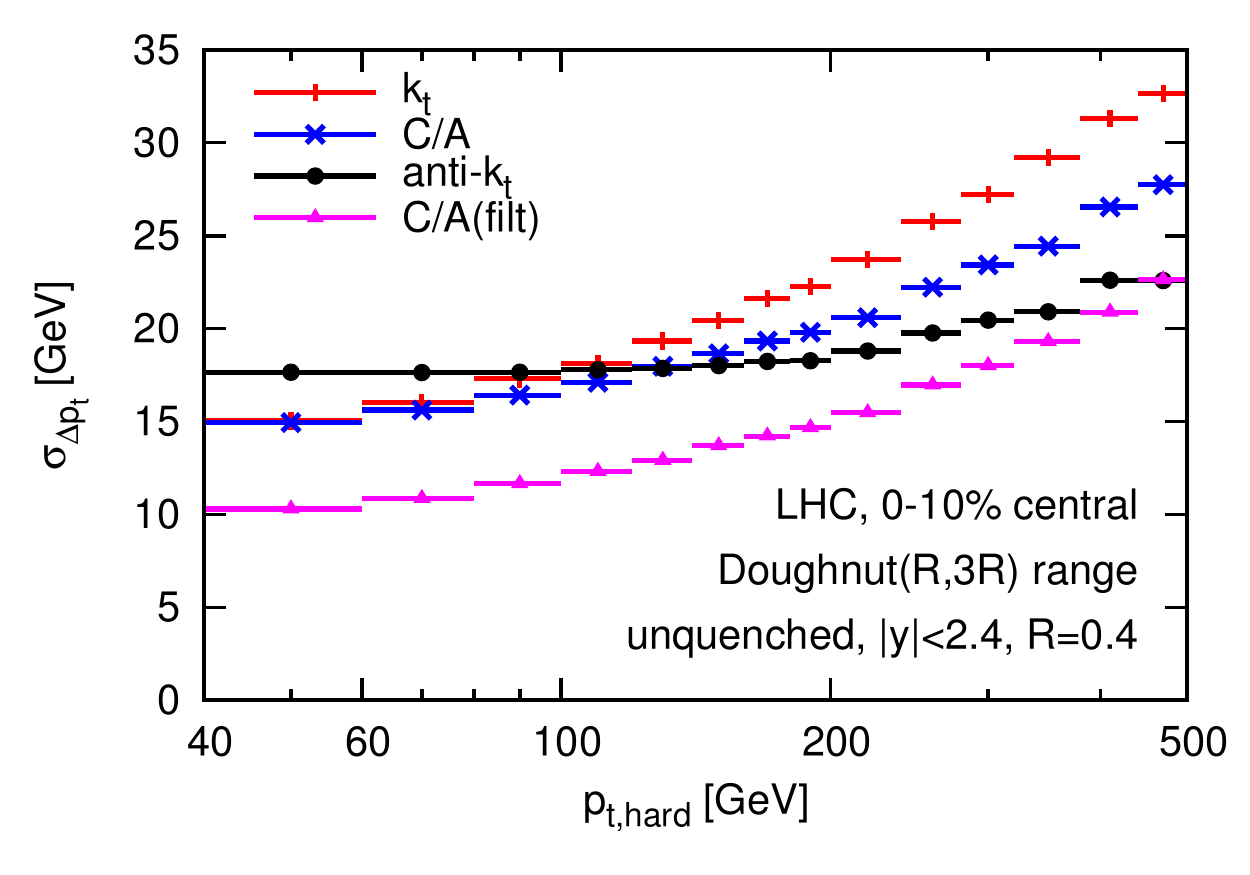}
\end{center}
\caption{Dispersion $\sigma_{\Delta p_t}$ for the LHC kinematics.}\label{fig:disp}
\end{figure}

Finally, we turn to the dispersion $\sigma_{\Delta p_t}$. The observed
dispersions for the LHC are shown on Fig. \ref{fig:disp}. While
anti-$k_t$ was the more robust in terms of average shift, when we
consider the dispersion, C/A(filt) shows better performances. What we
really see here is precisely the effect of filtering which practically
reduces the area of the jet, thus reducing the dispersion which is
proportional to $\sqrt{A_{\rm jet}}$. With our choice of parameters,
the area after filtering is about $1/2$ of the original C/A-jet area
and we may thus expect a decrease of a factor $1/\sqrt{2}$ in the
dispersion, which is consistent with what is observed on the figure at
low $p_t$.

\paragraph{Additional remarks} 

To conclude this Section, we want to make a few remarks on other tests
that can be performed (see \cite{hipaper} for more details).

First, to test the robustness of our results, we have checked the
effect of quenching on the embedded hard events. This can been done
{\em e.g.} by running PyQuen \cite{pyquen} on the Pythia jets. We have
noticed a negligible effect on the results obtained with the
anti-$k_t$ algorithm and an effect smaller than 2\% for the C/A(filt)
case.

Then, we can investigate if our conclusions hold for more peripheral
collisions, where flow effects may increase the fluctuations. For the
average quantities presented here, we have checked that our
conclusions remain unchanged in the case of non-central
collisions. However, it is interesting to mention that if we study the
average shift $\avg{\Delta p_t}$ as a function of the azimuthal angle between
the jet and the reaction plane, we see oscillations of a few GeV. The
use of a range with a finite azimuthal coverage, like the
DoughnutRange, reduces these oscillations compared to a StripRange,
but further developments would be needed to reduce them completely.

\section{Conclusions}\label{sec:ccl}

To conclude, we have studied the possibility to use the jet-area-based
subtraction method initially proposed in \cite{bkgestim} to
reconstruct full jets in heavy-ion collisions. 

We have found that the use of local ranges to reduce the
non-uniformities of the background (like its rapidity dependence)
improves the background subtraction. It would be interesting to see if
using different ranges (bearing in mind that they should contain at
least $\sim$9 jets to give a reliable estimate of $\rho$) could help
obtaining an unceratinty on the estimation of $\rho$.

We have studied subtraction performances by embedding Pythia events in
Hydjet heavy-ion events and measured for the two hardest jets in each
event $\Delta p_t$, the transverse momentum difference between the
original (subtracted) hard jet and the embedded/full jet after
subtraction. While knowledge of the full $\Delta p_t$ distribution is
relevant, {\em e.g.} for unfolding purposes \cite{jacobs}, we have
chosen to focus only on its average and dispersion.

For the average shift $\avg{\Delta p_t}$, we have observed that the
anti-$k_t$ (because of its rigidity) and the C/A algorithm
supplemented with a filter give a result close to zero, though one has
to notice that for the latter, this is the result of a cancellation
between two effects.

In terms of dispersion the C/A(filt) algorithm, which has a smaller
sensitivity to the UE, as a consequence of its reduced area, shows
better performances than the other algorithms.

Finally, we have noticed (see \cite{hipaper} for more details) that
these results also hold when quenched jets are embedded or when one
considers more peripheral collisions. In this last case, there is a
residual shift when one considers the shift as a function of the
azimuthal angle between the jet and the reaction plane that is left
for further studies. We also note that, though the C/A filt algorithm
has a smaller dispersion, it may be more affected by quenching than
the anti-$k_t$ algorithm ({\em e.g.} because of jet broadening) and it
may thus be helpful to consider both to obtain the most complete and
valuable information.

\paragraph{Acknowledgements} I am grateful to the organisers of the
Hard Probes conference for the opportunity to discuss the results
presented here. I also want to thank M.~Cacciari, J.~Rojo and G.~Salam
which have contributed to the physics results presented in these
proceedings.

 
\bibliographystyle{elsarticle-num}
\bibliography{gsoyez}


\end{document}